\documentclass[aps,prb,twocolumn,showpacs,superscriptaddress,floatfix,10pt]{revtex4-1}

\usepackage{graphicx}
\usepackage{amsmath}
\usepackage{epsfig}
\usepackage{helvet}
\usepackage{amssymb}
\usepackage{framed}
\usepackage{enumitem}
\usepackage[pdftex,plainpages=false,bookmarksnumbered=true,hypertexnames=false,colorlinks=true,linkcolor=blue,urlcolor=blue,citecolor=blue,anchorcolor=green]{hyperref}

\newcommand{\bra}[1]{\mbox{$\langle #1 |$}}
\newcommand{\ket}[1]{\mbox{$| #1 \rangle$}}
\newcommand{\braket}[2]{\mbox{$\langle #1 | #2 \rangle$}}

\newcommand{\frw}[1]{$\overset{\lower0.5em\hbox{$\smash{\scriptscriptstyle\smile}$}} #1$}

\def\T{{\mathcal T}}

\def\L{{\mathcal L}}

\def\C{{\mathcal C}}

\def\Z{{\mathcal Z}}

\begin{document}

\title{Number-State Preserving Tensor Networks as Classifiers for Supervised Learning}
\author{Glen Evenbly}
\affiliation{School of Physics, Georgia Institute of Technology, Atlanta, GA 30332, USA}
\email{gevenbly3@gatech.edu}
\date{\today}

\begin{abstract}
We propose a restricted class of tensor network state, built from number-state preserving tensors, for supervised learning tasks. This class of tensor network is argued to be a natural choice for classifiers as (i) they map classical data to classical data, and thus preserve the interpretability of data under tensor transformations, (ii) they can be efficiently trained to maximize their scalar product against classical data sets, and (iii) they seem to be as powerful as generic (unrestricted) tensor networks in this task. Our proposal is demonstrated using a variety of benchmark classification problems, where number-state preserving versions of commonly used networks (including MPS, TTN and MERA) are trained as effective classifiers. This work opens the path for powerful tensor network methods such as MERA, which were previously computationally intractable as classifiers, to be employed for difficult tasks such as image recognition.
\end{abstract}

%\pacs{05.30.-d, 02.70.-c, 03.67.Mn, 75.10.Jm}
\maketitle

\section{Introduction} \label{sect:intro}
Ideas and methods from the field of machine learning are currently having a significant impact in many areas of physics research \cite{Carleo19}. Machine learning offers powerful new tools for classifying phases of matter \cite{Carras17,Bro17,Chng17,Hue18,Liu18,Cana19}, for processing experimental results \cite{Tor18,Carras19}, and for modeling quantum many-body systems \cite{Tor16,Carleo17,Choo19}, to name but a few of the plethora of applications. With this crossing of fields has come the intriguing realization that the neural networks\cite{Hass95,Scha97} used in machine learning share extensive similarities with the tensor networks\cite{Orus14} used in modeling quantum many-body systems\cite{Levi18}. These connections are perhaps not so surprising since both types of network have the primary function of encoding large sets of correlated data: neural networks encode ensembles of training data, while tensor networks encode superpositions of quantum states. Currently there is great interest in exploring the potential applications of this relation, both from the directions of (i) using ideas from neural networks and machine learning to improve methods for modeling quantum wave-functions\cite{Huang17,Deng17,Glas18,Cai18} and (ii) examining tensor networks as a new approach for tasks in machine learning\cite{Miles16,Cohen17,Han17,Cich17, Liu17,Hallam17,Miles18,Liu18b,Hugg18, Grant18,Glas18b}.

In this manuscript we focus on the second direction (ii), and explore the use of tensor networks as classifiers for supervised learning problems. Research in this area has already produced encouraging early results, with examples where tensor networks have been trained to produce relatively competitive classifiers in both supervised and unsupervised learning tasks\cite{Miles16, Liu17,Hallam17,Miles18,Grant18,Glas18b}. However there are some significant issues with respect to the use of tensor networks as classifiers. One such issue is that of \emph{interpretability}. Usually, when applying a tensor network as a classifier, each sample from the (classical) dataset is associated to a product state. However, under generic tensor transformations, product states can be mapped to entangled quantum states, which can no longer be re-interpreted classically. One can understand this as a problem of generic tensor networks being overly-broad when used as classifiers: they are designed to carry information about phases and/or signs between superposition states, which are \emph{necessary} for describing wave-functions but seem to be \emph{extraneous} from the perspective of characterizing classical datasets. A second issue is that of computational efficiency. Most previous studies have utilized only relatively simple classes of tensor networks, such as matrix product states\cite{MPS1,MPS2} (MPS) and tree tensor networks\cite{TTN1,TTN2} (TTN), as classifiers. The more formidable weapons in the arsenal of tensor networks, such as the multi-scale entanglement renormalization ansatz\cite{MERA1,MERA2,MERA3,MERA4} (MERA), which are seen as the direct analogues to the high successful convolutional neural networks\cite{CNN1,CNN2,CNN3} (CNNs), have yet to be deployed in earnest for challenging problems. The primary reason being that, in order for a tensor network to be of use as a classifier, ones needs to be able to compute scalar products between the network and product states (representing the training data); this can be done efficiently for simple networks such as MPS and TTN, but is generally computationally intractable for more sophisticated networks like MERA. 

The main motivation for this manuscript is to help resolve the two issues discussed above. In particular, we propose to use networks built from a restricted class of tensor, those which act to preserve number-states, as classifiers for supervised learning tasks. Such number-state preserving networks automatically resolve the issue of interpretability, provided that each sample of the training data is encoded as a number state. Moreover, the restriction to number-state preserving tensors endows networks with a causal cone structure when contracted against number states, similar to the causal cone structure present in isometric networks when contracted against themselves. This property allows for a broad class of number-state preserving networks, including versions of MERA, to be efficiently trained as classifiers for supervised learning problems. Furthermore, we demonstrate numerically that networks built from this restricted class of number-state preserving tensor perform well for several example classification problems. The above considerations indicate that number-state preserving tensors are a natural restriction to impose when applying tensor methods to learn from sets of classical data.

This manuscript is organized as follows. Firstly in Sect. \ref{sect:preserve}, we characterize number-state preserving tensors and some of their properties, then in Sect. \ref{sect:supervised} we formulate how problems in supervised learning can be approached using tensor networks. In Sect. \ref{sect:updates} we propose an algorithm for training number-state preserving tensor networks to correctly classify a labeled dataset, while Sect. \ref{sect:enviro} we describe how single tensor environments can be efficiently evaluated, a key ingredient in the proposed training algorithm. Benchmark numerical results for number-state preserving versions of MPS, TTN and MERA applied to example classification problems are presented in Sect. \ref{sect:bench}, and conclusions are presented in Sect. \ref{sect:conclusion}.

%%%%%%%%%%%%%%%%%%%%%%%%%%%%%%%%%%%%%%
%%%%%%%%%%%%%%%%%%%%%%%%%%%%%%%%%%%%%%
\begin{figure}[!t!b]
\begin{center}
\includegraphics[width=8.5cm]{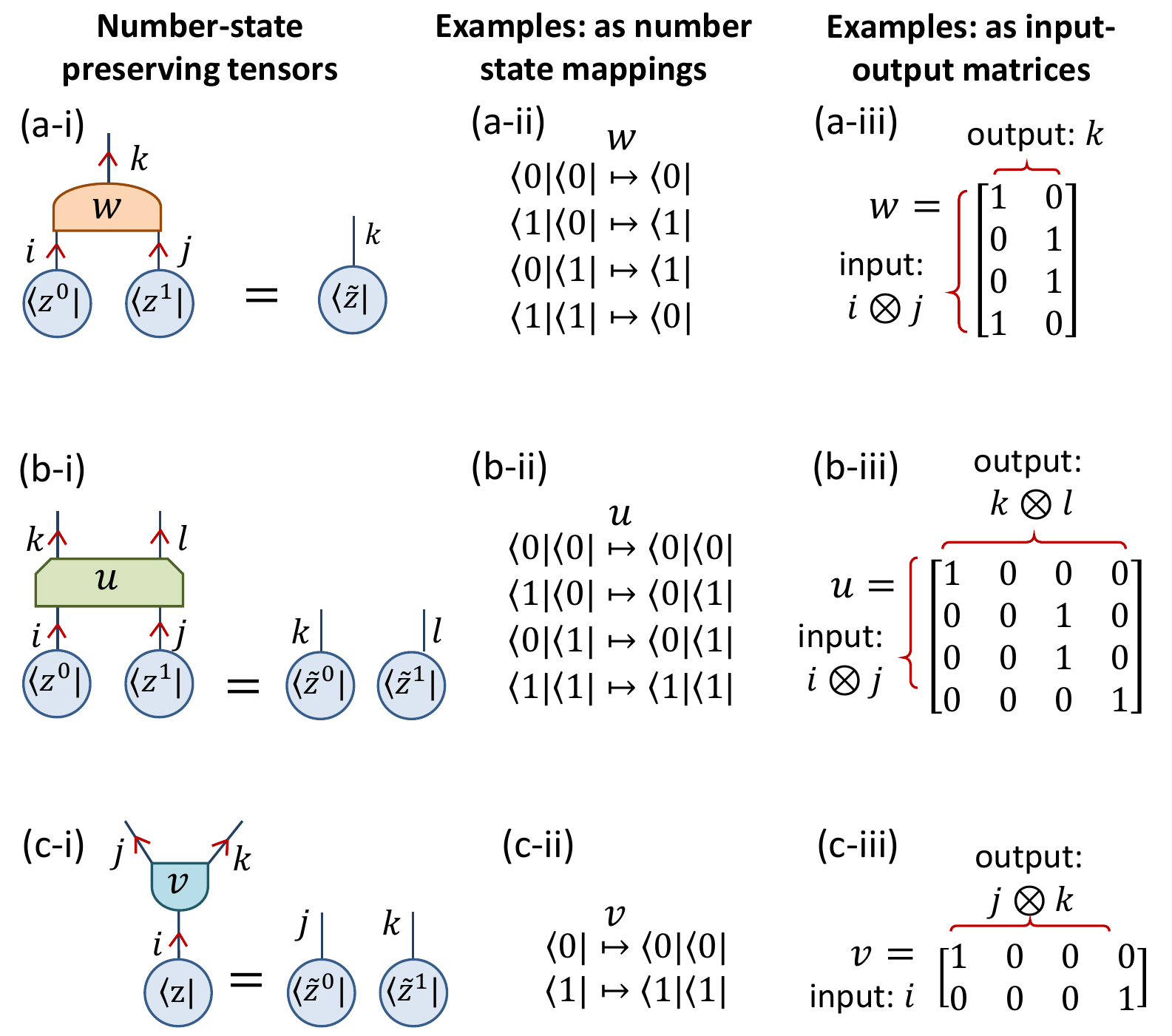}
\caption{(a) An example of a number-state preserving tensor that $w$ maps a number state $\bra{z^0}\bra{z^1}$ on its input indices to a number state $\bra{\tilde z}$ on its output index. The tensor $w$ can be equivalently represented as (ii) an explicit mapping between number states or (iii) as a matrix (after forming the product of input indices). (b) An example of a number-state preserving tensor $u$ between two input and two output indices. (c) An example of a number-state preserving tensor $v$ between one input and two output indices. Note that the three examples of  number-state preserving tensors from (a-c) are also \emph{unital}, in that all of their non-zero entries are the unit element.} 
\label{fig:NStensors}
\end{center}
\end{figure}
%%%%%%%%%%%%%%%%%%%%%%%%%%%%%%%%%%%%%%%%
%%%%%%%%%%%%%%%%%%%%%%%%%%%%%%%%%%%%%%%%

\section{Number-state preserving networks} \label{sect:preserve}
Let $\L$ be a lattice of sites, with each site described by a local Hilbert space of some dimension $d$. We label the basis states for each site by integers, $\ket{z} \in \{\ket{0}, \ket{1},\ldots, \ket{d-1} \}$, which are interpreted as particle number and are represented as unit vectors,
\begin{equation}
\left| 0 \right\rangle  = \left[ {\begin{array}{*{20}{c}}
1\\
0\\
0\\
0\\
 \vdots 
\end{array}} \right],\; \; \left| 1 \right\rangle  = \left[ {\begin{array}{*{20}{c}}
0\\
1\\
0\\
0\\
 \vdots 
\end{array}} \right],\; \; \left| 2 \right\rangle  = \left[ {\begin{array}{*{20}{c}}
0\\
0\\
1\\
0\\
 \vdots 
\end{array}} \right],\ldots \label{eq:UnitVec}
\end{equation}
A number state $\ket{{\Z^{\L}}}$ (or, equivalently, a Fock state) on lattice $\L$ is a product state with well-defined particle number, 
\begin{equation}
\left| {{\Z^{\L}}} \right\rangle  = \ket{z^0} \ket{z^1} \ket{z^2}  \ldots, \label{eq:number}
\end{equation}
where superscripts are here used to denote lattice position. Alternatively, if one is thinking in terms of spin degrees of freedom, a number state can be defined as a product state with a well-defined $z$-component of spin.

\begin{figure}[!t!b]
\begin{center}
\includegraphics[width=8.5cm]{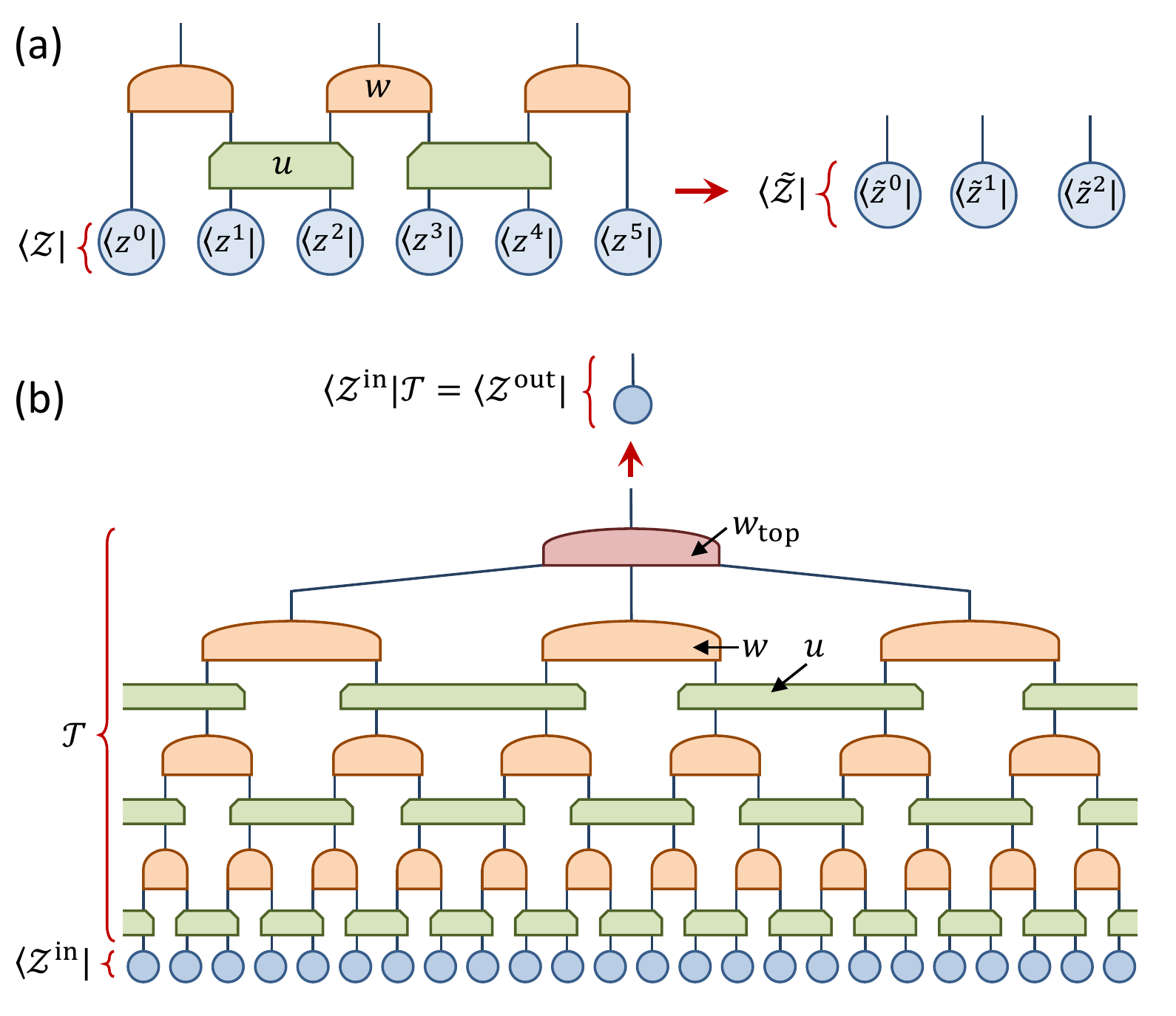}
\caption{(a) A number-state preserving network is formed through composition of number-state preserving tensors $u$ and $w$, which maps input number state $\bra{\mathcal{Z}}$ to output $\bra{\mathcal{\tilde Z}}$. (b) A binary MERA tensor network $\mathcal{T}$, assumed to be composed of number-state preserving tensors, maps an input number state $\bra{\Z^\textrm{in}}$ on a lattice of 24 sites to an output number state, $\bra{\Z^\textrm{in}} \T \mapsto \bra{\Z^\textrm{out}}$, on a single site.} 
\label{fig:NSnetwork}
\end{center}
\end{figure}

We now turn our considerations to transformations of number-states implemented by certain types of \emph{oriented} tensor: these are tensors where each index has been fixed as either \emph{incoming} or \emph{outgoing}. Any oriented tensor can be interpreted as a mapping between states defined on an input lattice $\L$, whose sites match the incoming tensor indices, to states on an output lattice $\L'$, whose sites match the outgoing tensor indices. We define an oriented tensor as \emph{number-state preserving} if it maps any number state defined on $\L$ to another number state on $\L'$. Several examples of number-state preserving tensors are given in Fig. \ref{fig:NStensors}. Let $u_{ij}^{kl}$ be a four index tensor, with subscripts denoting incoming indices and superscripts denoting outgoing indices, as depicted in Fig. \ref{fig:NStensors}(b). Consider the reshape of $u$ into an input-output matrix, i.e. where the rows of the matrix enumerate over the tensor product $(i\otimes j)$ of incoming indices and columns enumerate over the tensor product of the outgoing indices $(k\otimes l)$. It is easily understood that the property of $u$ being number-state preserving is equivalent to the property that each row of the corresponding input-output matrix must have \emph{at most} a single non-zero entry. Note that we also include in the definition of number-state preserving tensors those where the input-output matrix has rows with only zero entries; equivalently these are tensors which can map some number states to the null (or norm-zero) state. An important property of number-state preserving tensors is that networks formed from their composition, where outputs from one tensor are properly matched with inputs to other tensors, are also number-state preserving, as depicted in Fig. \ref{fig:NSnetwork}(a). This allows us to form number-state preserving versions of commonly used tensor networks, such as MERA, as shown in Fig. \ref{fig:NSnetwork}(b). However, it is vital to realize that number-state preserving tensors do not necessarily remain number-state preserving if the orientation of their indices is reversed (i.e. the incoming and outgoing indices are switched); thus number-state preserving networks can still generate interesting superpositions and entangled states when `run' in reverse.

%%%%%%%%%%%%%%%%%%%%%%%%%%%%%%%%%%%%%%
%%%%%%%%%%%%%%%%%%%%%%%%%%%%%%%%%%%%%%
%%%%%%%%%%%%%%%%%%%%%%%%%%%%%%%%%%%%%%%%
%%%%%%%%%%%%%%%%%%%%%%%%%%%%%%%%%%%%%%%%

For the main text of this paper we shall further restrict our consideration to \emph{unital} number-state preserving tensors, where each tensor entry must be either a zero or a one, and each row of the corresponding input-output matrix is required to have a single non-zero entry. Note that this class of tensor maps incoming number-states to outgoing number-states of the \emph{same} normalization and phase. The restriction to unital tensors will be useful in simplifying their application to supervised learning problems, although the formalism and optimization algorithms that we present are still general for all number-state preserving networks. There are many reasons why one may also wish to consider networks comprised of non-unital number-state preserving tensors, where entries can take any real or complex value, and thus change the normalization of states and introduce phases; the interested reader is directed to Sect. \ref{sect:A} of the Appendix for further discussion. 

Given that number-state preserving networks represent a severely restricted class of tensor network states it may be interesting to consider how much of their power has been lost, for instance, in describing ground states of quantum many-body systems. Although this remains to be explored, it seems likely that majority of many-body systems will not have ground-states that can be well-approximated by number-state preserving tensor networks. However, there does exist several examples of non-trivial quantum many-body systems related to Motzkin paths \cite{Motz1}, whose ground states possess interesting entanglement and yet can be exactly represented by number-state preserving networks \cite{Motz2,Motz3}. Investigation of the ability of number-state preserving networks to describe general quantum ground states remains an intriguing direction for future research. 

%%%%%%%%%%%%%%%%%%%%%%%%%%%%%%%%%%%%%%
%%%%%%%%%%%%%%%%%%%%%%%%%%%%%%%%%%%%%%
\begin{figure}[!t!b]
\begin{center}
\includegraphics[width=8.5cm]{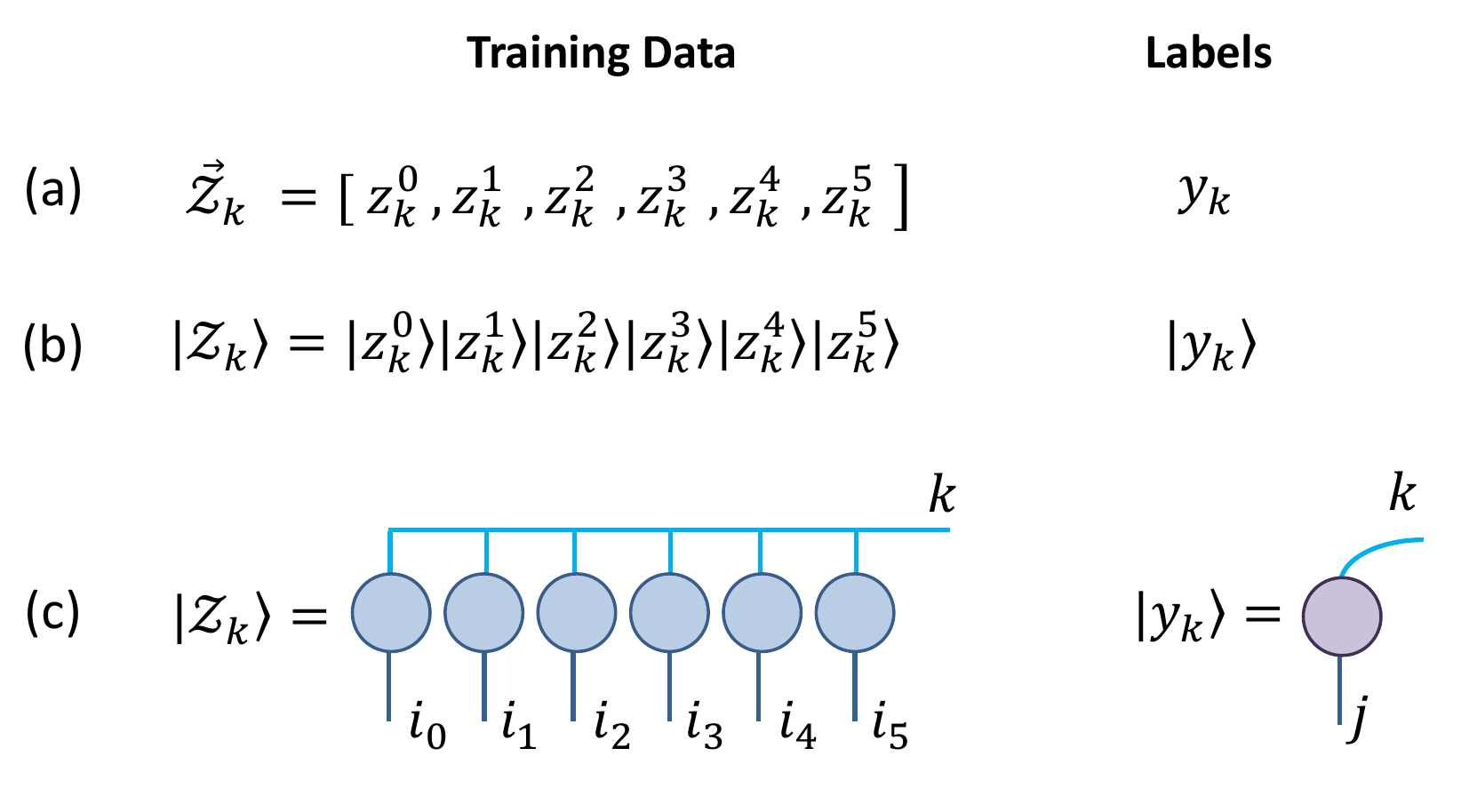}
\caption{(a) The $k^\textrm{th}$ training sample $\vec{\Z}_k$ is given as a length $N$ vector of integers $z_k$ (modulo some specified base $d$), and is accompanied by label $y_k$. (b) The training sample $\vec{\Z}_k$ can alternatively be expressed as a unit vector $\ket{{\Z}_k}$ in the tensor product space of dimension $d^N$ formed from mapping each base-$d$ integer to a number-state $\ket{z_k}$, see Eq. \ref{eq:UnitVec}. (c) Diagrammatic tensor representation of training sample $\ket{{\Z}_k}$.} 
\label{fig:MLdata}
\end{center}
\end{figure}
%%%%%%%%%%%%%%%%%%%%%%%%%%%%%%%%%%%%%%%%
%%%%%%%%%%%%%%%%%%%%%%%%%%%%%%%%%%%%%%%%

\section{Supervised learning in a tensor product space} \label{sect:supervised}
In this section we discuss how the task of supervised learning can be formulated in terms of tensor networks. We consider problems where each training sample $\vec \Z$ is represented as a length $N$ vector, with the $i^\textrm{th}$ component $z^i$ an element of $\mathbb{Z}_d$ (the set of integers modulo $d$), i.e. such that
\begin{equation}
{\vec Z_k} = \left[ {z_k^0,z_k^1,z_k^2, \ldots ,z_k^{N - 1}} \right], \label{eq:data}
\end{equation}
where $k$ is a label over the set of training samples. Every training sample is assumed to be paired with a corresponding label $y \in \mathbb{Z}_c$, where $c$ represents the number of distinct categories for the classification problem. The goal of the supervised learning problem is to construct a function $f$ that maps each sample of the training set to its correct label,
\begin{equation}
f:{\vec Z_k} \mapsto {y_k}. \label{eq:function}
\end{equation}
Although classifiers based on linear functions $f$ have some considerable utility \cite{Kern1}, many non-trivial classification problems require non-linear functions $f$ in order to achieve good accuracy.

We now describe how a tensor network can be implemented as the classifying function in Eq. \ref{eq:function}. At this point, one could be tempted to believe that tensor networks would have limited utility as classifiers as, given that tensors simply are extensions of matrices to higher dimensions, they are inherently \emph{linear} constructs. However, in order to recast the supervised learning problem into a problem amenable to tensor networks, we first (non-linearly) embed the training data into a higher dimensional space, similar to a kernel method \cite{Kern2}. By using an appropriate non-linear embedding, a linear classifier acting the higher dimension space can reproduce the classifying power of non-linear functions in the original space; thus it remains possible that tensor network approaches could be competitive with classifiers based on (non-linear) neural networks. Indeed, as will be argued later in this manuscript, it can be understood that a tensor network of sufficiently large bond dimension $\chi$ can, in principle, obtain perfect accuracy for any training set of a supervised learning problem as formulated above. 

Let us recast each training sample ${\vec \Z}_k$ as a number state, denoted $\ket{{ \Z}_k}$, defined in a vector space of total dimension $d^N$. Specifically, we associate each integer $z \in \mathbb{Z}_d$ with a number state $\ket{z}$ in a $d$-dimensional Hilbert space, represented as per Eq. \ref{eq:UnitVec},
such that the full state vector $\ket{{ \Z}_k}$ is given as the tensor product of the single site states,
\begin{equation}
\left| {{\Z_k}} \right\rangle  = \left| {z_k^0} \right\rangle \left| {z_k^1} \right\rangle \left| {z_k^2} \right\rangle  \ldots \left| {z_k^{N - 1}} \right\rangle.  \label{eq:FullVec}
\end{equation}
Similarly the data labels $y_k$ are recast as number states $\ket{y_k}$ in a $c$-dimensional space. The diagrammatic tensor notation for these states is presented in Fig. \ref{fig:MLdata}. Given this embedding of our training data, a classifier can be represented as tensor network $\T$ that maps states $\bra{\Z_k^\textrm{in}}$ from the lattice of $N$ sites of dimension $d$ to states $\bra{\Z_k^\textrm{out}}$ on a single site of dimension $c$,
\begin{equation}
\bra{\Z_k^\textrm{in}} \T = \bra{\Z_k^\textrm{out}}, \label{eq:Tmap}
\end{equation}
see also Fig. \ref{fig:NSnetwork}(b) for an explicit example.

\begin{figure}[!t!b]
\begin{center}
\includegraphics[width=8.5cm]{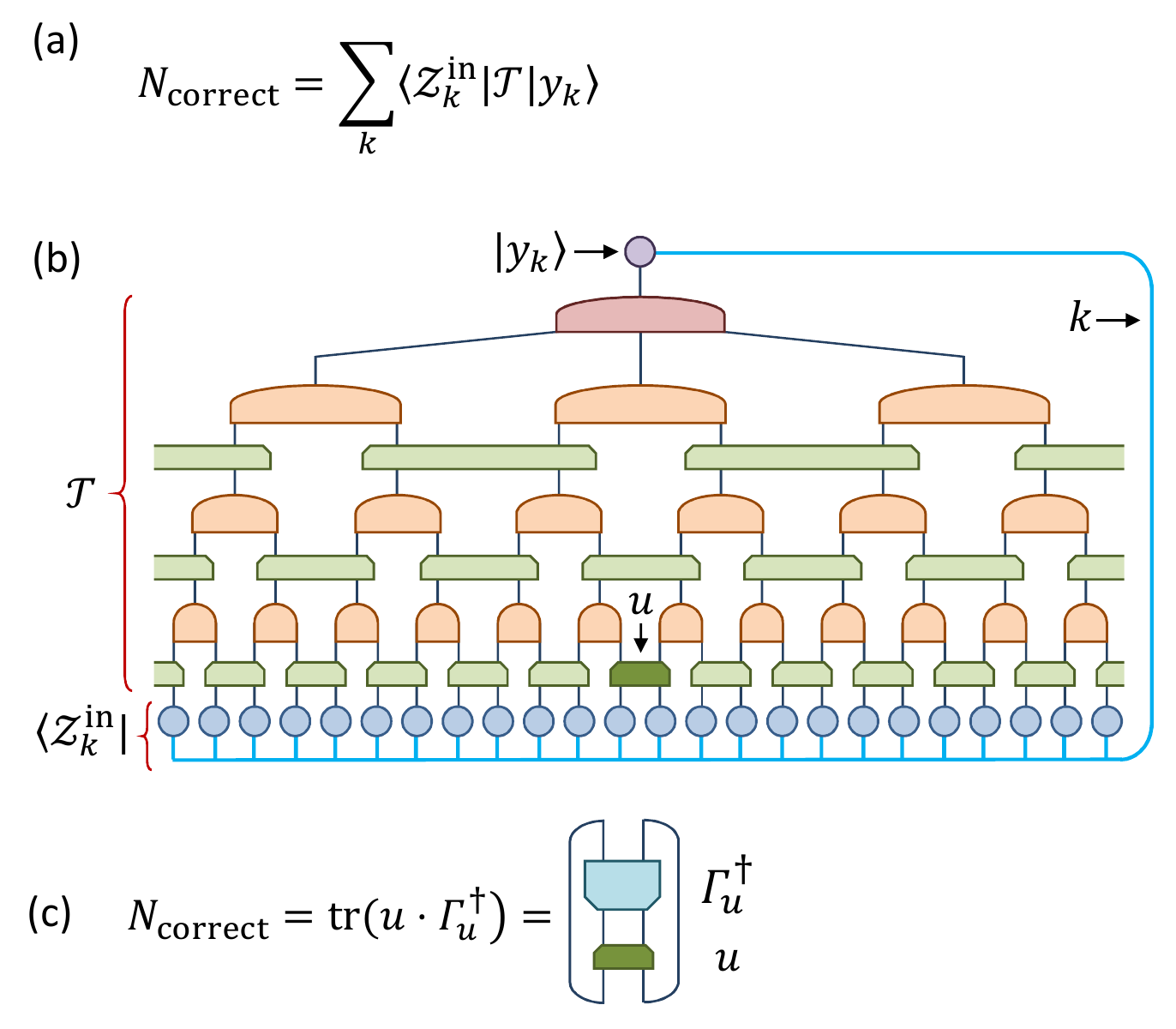}
\caption{(a) The total number correctly classified samples $N_\textrm{correct}$ is given as the inner product of the labels $\ket{y_k}$ against the network $\T$ applied to the training data $\bra{\Z_k^\textrm{in}}$, summing over all training samples $k$. (b) Diagrammatic representation of the equation from (a) which evaluates to $N_\textrm{correct}$. (c) For any chosen tensor, such as the shaded tensor $u$ in (b), the network for $N_\textrm{correct}$ can be factorized into a product of the tensor with its environment $\Gamma_u$, formed from contracting the entirety of the network sans $u$. The environment $\Gamma_u$ allows the optimal tensor $u$ that maximizes $N_\textrm{correct}$ (with the other tensors in $\T$ held fixed) to be identified.} 
\label{fig:Ncorr}
\end{center}
\end{figure}

In general, the accuracy of $\T$ as a classifier could be quantified by evaluating the scalar products of the output states with the label states, $\braket{\Z_k^\textrm{out}}{y_k}$, where a large scalar product would indicate good classification. However, in the particular case of \emph{unital} number-preserving networks $\T$, the norm of states is preserved such that all scalar products $\braket{\Z_k^\textrm{out}}{y_k}$ either evaluate to unity (indicating correct classification of the data sample with label $y_k$) or to zero (indicating incorrect classification of the data sample). Thus, the number of correctly classified samples $N_\textrm{correct}$ simply evaluates as the sum over all the scalar products,
\begin{equation}
{N_\textrm{correct}} = \sum\limits_k {\left\langle {\Z_k^\textrm{in}} \right|\T\left| {{y_k}} \right\rangle }. \label{eq:Ncorr}
\end{equation}
The diagrammatic tensor notation for Eq. \ref{eq:Ncorr}, in the particular case that $\T$ is a binary MERA, is presented in Fig. \ref{fig:Ncorr}(b). It follows we should use Eq. \ref{eq:Ncorr} as the \emph{cost function} for training the tensor network $\T$ for the supervised learning problem: the tensors contained within $\T$ should be optimized as to maximize ${N_\textrm{correct}}$. Methods for achieving this are discussed in the following section of this manuscript.

Before moving on, we remark that the formalism we described (or similar formalisms consider previously \cite{Miles16,Han17,Hallam17,Miles18,Liu18b,Glas18b}) for addressing supervised learning problems using tensor networks could, in principle, employ arbitrary tensor networks $\T$ as classifiers (not only those built from number-state preserving tensor networks). However, it is only for certain types of network, such as MPS and TTN, that scalar products of the form ${\left\langle {{\Z_k^\textrm{in}}} \right|\T\left| y_k \right\rangle }$ can be efficiently evaluated. The cost of (exactly) evaluating the overlap of a product state with a more sophisticated tensor network state, such as a MERA, typically does not scale efficiently with system size. Thus, one would expect that a general MERA network would only be computationally feasible as a classifier for problems with a small number of sites (or variables). In contrast the output state $\bra{\Z_k^\textrm{out}}$ of Eq. \ref{eq:Tmap} can be efficiently evaluated for \emph{any} number-state preserving tensor network, with cost that scales only linearly in the number of tensors in $\T$. Nonetheless, the result that a scalar product ${\left\langle {{\Z_k^\textrm{in}}} \right|\T\left| y_k \right\rangle }$ is efficient to evaluate does not in itself imply that the network $\T$ can be efficiently trained. In Sect.\ref{sect:enviro} we formulate additional requirements for network $\T$ that are sufficient to allow for efficient training. 

%%%%%%%%%%%%%%%%%%%%%%%%%%%%%%%%%%%%%%
%%%%%%%%%%%%%%%%%%%%%%%%%%%%%%%%%%%%%%
%%%%%%%%%%%%%%%%%%%%%%%%%%%%%%%%%%%%%%%%
%%%%%%%%%%%%%%%%%%%%%%%%%%%%%%%%%%%%%%%%

\section{Single tensor updates} \label{sect:updates}
In this section we propose a method to optimize the tensors of a network $\T$ to maximize the number ${N_\textrm{correct}}$ of correctly identified training samples in a supervised learning problem, as formulated in Eq. \ref{eq:Ncorr}. We follow the same strategy of \emph{single tensor updates} developed in the context optimizing MERA \cite{Alg1}, where only a single tensor in the network is changed at any time while all other tensors in the network are held fixed. These single tensor updates can then be organized into `sweeps', in which all tensors in the network are optimized in turn, and the sweeps iterated until the entire network is sufficiently converged.
 
Key to this optimization strategy is the notion of a \emph{tensor environment}, which can be understood as the derivative of the network with respect to a single tensor. Specifically, given a network that evaluates to a scalar such as that from Fig. \ref{fig:Ncorr}(b), the environment $\Gamma_u$ of a tensor $u$ results from contracting the entire network sans the particular tensor $u$ under consideration. It follows that the number of correctly classified samples ${N_\textrm{correct}}$ from Eq. \ref{eq:Ncorr} can always be expressed as the scalar product of a tensor $u \in \T$ with its environment $\Gamma_u$,
\begin{equation}
{N_\textrm{correct}} = \textrm{tr}(u \cdot \Gamma _u^\dag ), \label{eq:update}
\end{equation}
where, for notational simplicity, we have recast $u$ and $\Gamma_u$ into input-output matrices, see Fig. \ref{fig:Ncorr}(c). We relegate a description of the general method for computing environments $\Gamma_u$ to Sect.\ref{sect:enviro} of the manuscript, and proceed here assuming $\Gamma _u$ is already known. 

Let us now turn to the problem of finding the optimal number-state preserving tensor $u_\textrm{opt.}$,
\begin{equation}
{u_\textrm{opt.}} \equiv \mathop \textrm{argmax }\limits_u  \Big[ \textrm{tr}\left( {u \cdot \Gamma _u^\dag } \right) \Big]
\end{equation}
which maximizes the number of correctly identified samples ${N_\textrm{correct}}$ of Eq. \ref{eq:update}, given a known environment $\Gamma _u$. Here it is easy to see that $u_\textrm{opt.}$ can be built by simply identifying the location of the maximal element in each row of $\Gamma _u$ and then placing the unit element at the corresponding location in each row of $u_\textrm{opt.}$, with all other entries zero. Note that if the maximal element in a row of $\Gamma _u$ is degenerate then $u_\textrm{opt.}$ is not uniquely defined; one can still obtain \emph{an} optimal solution by simply selecting one of the maximal elements in that row of $\Gamma _u$. Let us consider a concrete example: imagine we are updating a tensor $u$ with a $4\times 4$ input-output matrix of the form given in Fig. \ref{fig:NStensors}(b-iii), and assume that the environment has been evaluated as
\begin{equation}
\Gamma_u = \left[ {\begin{array}{*{20}{c}}
{10}&{12}&9&8\\
5&6&9&2\\
{21}&{18}&7&{22}\\
{12}&{15}&{13}&{14}
\end{array}} \right]. \label{eq:gamma}
\end{equation}
Then the (unital and number-state preserving) $4\times 4$ matrix $u_\textrm{opt.}$ that maximizes Eq. \ref{eq:update} is given as
\begin{equation}
u_\textrm{opt.} = \left[ {\begin{array}{*{20}{c}}
0&1&0&0\\
0&0&1&0\\
0&0&0&1\\
0&1&0&0
\end{array}} \right]. \label{eq:uoptimal}
\end{equation}
and the number of correctly classified training samples after this optimal update is given as ${N_\textrm{correct}}=(12+9+22+15) = 58$. Some remarks are in order regarding this optimization strategy. Firstly, we notice that unlike many commonly used algorithms for training neural networks, our approach is not based upon a gradient descent. Instead we can directly `hop' to the true maximum for any single tensor (given that the other tensors in the network are held remain fixed), provided the environment is exactly known. While this strategy has some advantages over gradient based methods with respect to avoiding local maxima, getting stuck in a solution that is not globally optimal can still remain a possibility depending on the problem until consideration.

We now discuss methods to introduce some randomness into the optimization, in order to reduce the possibility of getting trapped in a local maxima. One approach could be to employ a similar strategy as used in the stochastic gradient descent methods \cite{SGD1}, where randomness is introduced by using only select `batch' of training samples for each update. Instead, here we advocate a different strategy inspired by Monte Carlo methods \cite{Monte1} used in sampling many-body systems. Rather than updating to the optimal tensor $u_\textrm{opt.}$ at each step, we propose to allow updates to sub-optimal solutions of Eq. \ref{eq:update}, with a probability diminishes exponentially in relation to how far the solution is from the optimal solution. For this purpose we first introduce the difference matrix $\Omega$, given by subtracting from each row of $\Gamma$ the maximal element within the row,
\begin{equation}
{\Omega _{ij}} = {\Gamma _{ij}} - \mathop {\max }\limits_j \left( {{\Gamma _{ij}}} \right). \label{eq:omega}
\end{equation}
For the example environment $\Gamma_u$ given in Eq. \ref{eq:gamma} the corresponding difference matrix is
\begin{equation}
{\Omega} =  - \left[ {\begin{array}{*{20}{c}}
2&0&3&4\\
4&3&0&7\\
1&4&{15}&0\\
3&0&2&1
\end{array}} \right]. \label{eq:exomega}
\end{equation}
We then use the difference matrix to generate a matrix $p^\textrm{trans.}$ of transition probabilities, defined element-wise as 
\begin{equation}
p_{ij}^\textrm{trans.} = \frac{{\exp \left( {{\Omega _{ij}}/\alpha } \right)}}{{\sum\limits_j {\exp \left( {{\Omega _{ij}}/\alpha } \right)} }}\label{eq:prob}
%p_{ij}^\textrm{trans} = \exp ({\Omega _{ij}}/\alpha), \label{eq:prob}
\end{equation}
where $\alpha$ is a tunable parameter that sets the amount of randomness. For the example difference matrix $\Omega$ of Eq. \ref{eq:exomega} and setting $\alpha = 2$ we get the transition matrix
\begin{equation}
p^\textrm{trans.} = \left[ {\begin{array}{*{20}{c}}
{0.21}&{0.58}&{0.13}&{0.08}\\
{0.10}&{0.16}&{0.72}&{0.02}\\
{0.35}&{0.08}&{0.00}&{0.57}\\
{0.10}&{0.45}&{0.17}&{0.28}
\end{array}} \right].
\end{equation}
The transition matrix is then used to perform a stochastic update of the tensor $u$ under consideration: values in each row of $p^\textrm{trans.}$ set the probability for the unit element in the equivalent row of the updated $u$ to be placed at that particular location (note that Eq. \ref{eq:prob} has been defined such that each row of $p^\textrm{trans}$ sums to unit probability). Notice that in the limit $\alpha \rightarrow 0$ the matrix $p^\textrm{trans.}$ tends to $u_\textrm{opt.}$ (provided $\Gamma$ had no degeneracies in its maximal row values), since all non-optimal transitions are fully suppressed. Conversely, in limit $\alpha \rightarrow \infty$ all probabilities in $p^\textrm{trans.}$ tend to the same value, representing completely random transition probabilities.
 
%%%%%%%%%%%%%%%%%%%%%%%%%%%%%%%%%%%%%%
%%%%%%%%%%%%%%%%%%%%%%%%%%%%%%%%%%%%%%
\begin{figure}[!t!b]
\begin{center}
\includegraphics[width=8.5cm]{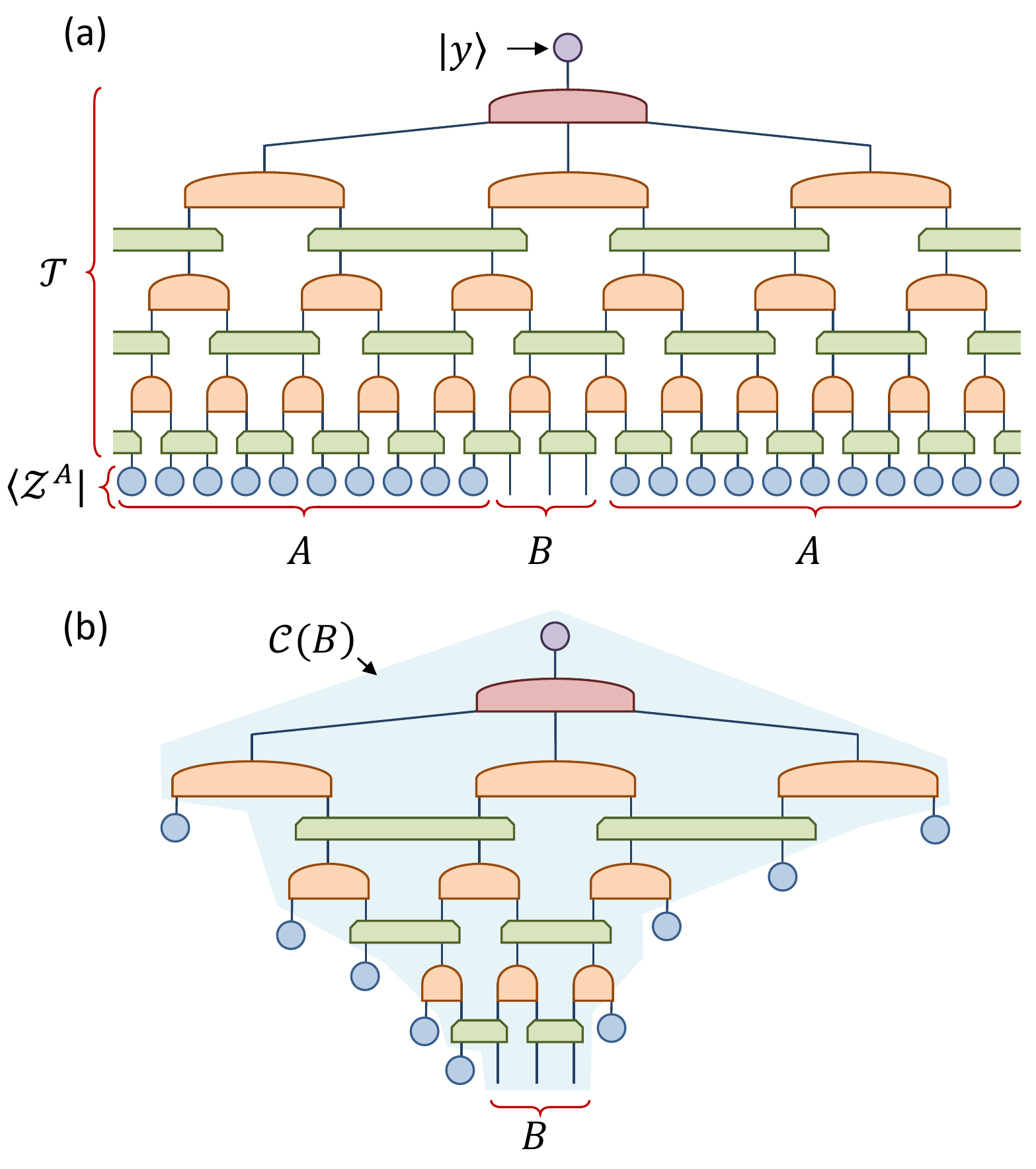}
\caption{(a) The network $\T$ with fixed output label $\ket{y}$ is applied to a number state $\bra{\Z^A}$ defined only on a sub-region $A$ of the initial lattice, with the state on the complimentary region $B$ left open. (b) The input number-state $\bra{\Z^A}$ is lifted through $\T$ as much as is possible by using the number-state mapping properties depicted in Fig. \ref{fig:NStensors}. The (configuration) causal cone $\C(B)$ associated to region $B$ describes the remaining set of tensors $\C \in \T$ after this lifting; this is equivalently the set of tensors whose output states can be affected by the choice of input state on region $B$. } 
\label{fig:Lifting}
\end{center}
\end{figure}
%%%%%%%%%%%%%%%%%%%%%%%%%%%%%%%%%%%%%%%%
%%%%%%%%%%%%%%%%%%%%%%%%%%%%%%%%%%%%%%%%
 
\section{Evaluation of tensor environments} \label{sect:enviro}
Here we describe evaluation of tensor environments, crucial to the optimization algorithm discussed in the previous section. For simplicity, we describe this evaluation assuming the tensor network $\T$ under consideration is a binary MERA, although the same methodology can be employed for arbitrary (number-state preserving) tensor networks. 

Rather than tackling the problem of computing tensor environments $\Gamma$ directly, we first introduce the concept of configuration spaces $\ket{\phi}$. Proper use of configuration spaces $\ket{\phi}$, which play an analogous role to the local reduced density matrices $\rho$ used to optimize tensor networks in the context of quantum many-body systems, will greatly simplify the subsequent evaluation of environments. Let us assume that the output index of the tensor network $\T$ under consideration has been fixed in some specified label state $\ket{y}$, and that the lattice on which it is defined has been partitioned into a region $A$ and its compliment $B$. Then, given a number state $\ket{\Z^A}$ on region $A$, we define the configuration space $\ket{\phi ^B}$ as 
\begin{equation}
\left| {{\phi ^B}} \right\rangle  = \sum\limits_{\textrm{configs: }\sigma } {\left| {\Z_\sigma ^B} \right\rangle }, \label{eq:config}
\end{equation}
where the sum runs over all valid configurations $\sigma$ of number states $\ket{\Z^B_\sigma}$ defined on region $B$ such that the combined number state ${\left| {{\Z^A}} \right\rangle \left| {\Z_\sigma ^B} \right\rangle }$ is classified by $\T$ into the correct category $\ket{y}$, i.e. such that
\begin{equation}
\Big( {\left\langle {{\Z^A}} \right|\left\langle {\Z_\sigma ^B} \right|} \Big) \T\left| y \right\rangle = 1. \label{eq:correct}
\end{equation}

An example of a network that could be contracted to evaluate a configuration space $\ket{{\phi ^B}}$ is depicted in Fig. \ref{fig:Lifting}(a). It is seen that this network can be simplified, as shown Fig. \ref{fig:Lifting}(b), by lifting the input number state $\ket{\Z^A}$ through tensors in $\T$ where-ever possible (i.e. where-ever a tensor has a number state available on all of its incoming indices), using the number-state preserving tensor properties as outlined in Fig. \ref{fig:NStensors}. It is convenient to define the \emph{configuration} causal cone $\C(B)$ associated to region $B$ as the set of tensors remaining in the network $\T$ after this simplification; equivalently $\C(B)$ can be defined as the set of tensors $\C \in \T$ whose output state can be affected by the choice of input state on region $B$. 

%%%%%%%%%%%%%%%%%%%%%%%%%%%%%%%%%%%%%%
%%%%%%%%%%%%%%%%%%%%%%%%%%%%%%%%%%%%%%
\begin{figure}[!t!b]
\begin{center}
\includegraphics[width=8.5cm]{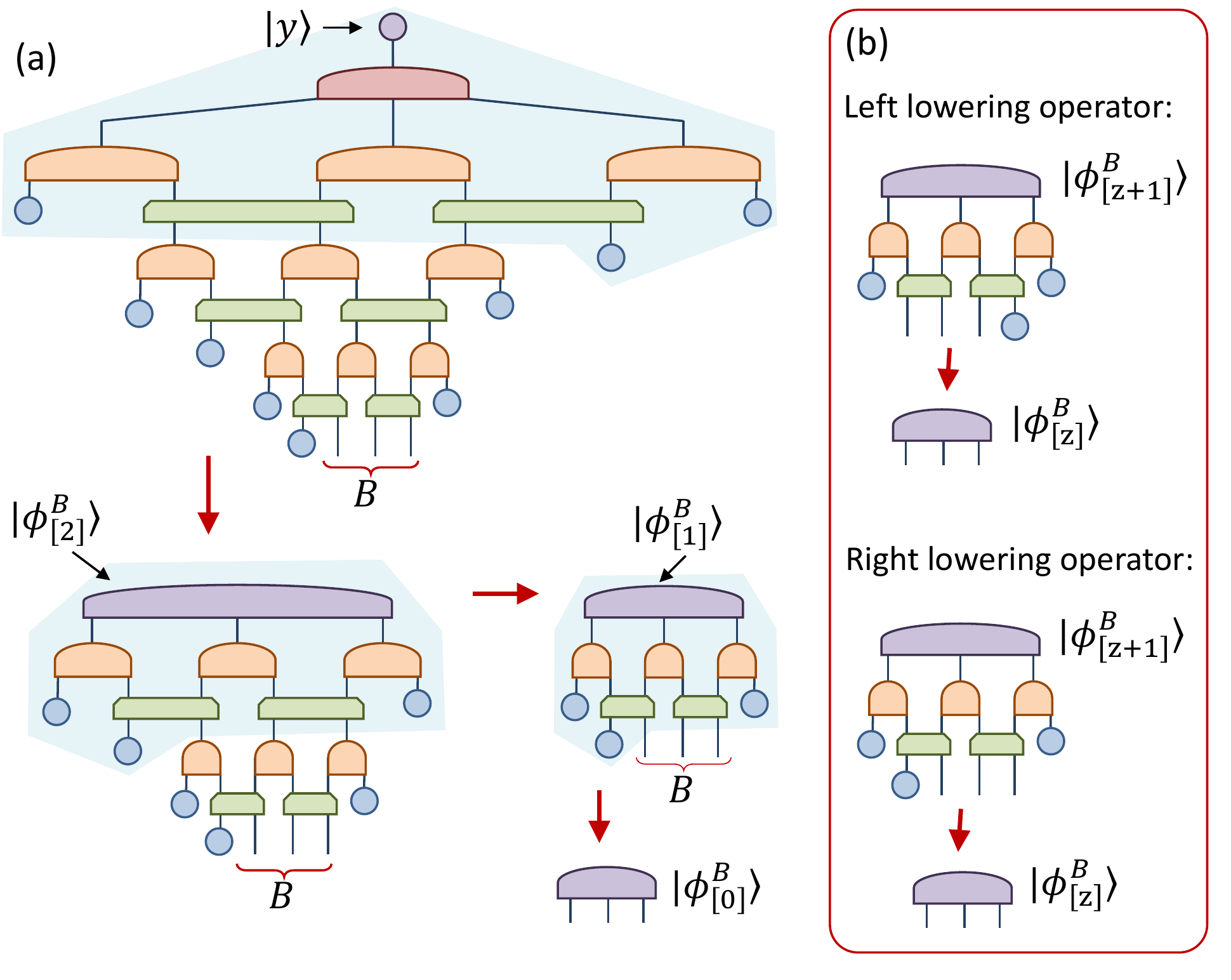}
\caption{(a) Sequence of contractions used to evaluate the configuration space $\ket{\phi^B_{[0]}}$ associated to region $B$, starting from the causal cone $\C(B)$ as depicted in Fig. \ref{fig:Lifting}(b). At each step in the evaluation the tensors in shaded region are contracted into a single tensor. (b) For any region $B$ of three contiguous sites on the initial lattice, the configuration space $\ket{\phi^B_{[0]}}$ can be evaluated using a composition of the left/right lowering operators.} 
\label{fig:CausalReduce}
\end{center}
\end{figure}
%%%%%%%%%%%%%%%%%%%%%%%%%%%%%%%%%%%%%%%%
%%%%%%%%%%%%%%%%%%%%%%%%%%%%%%%%%%%%%%%%

Notice that this configuration causal cone $\C(B)$ is precisely equivalent to the (standard) causal cone\cite{MERA1,Causal1} that would emerge from an isometric MERA for the same region $B$, defined as the set of tensors that can affect the local reduced density matrix $\rho_B$. However, the origins of these causal cones are drastically different: the causal cones in isometric MERA result arise due to the isometric constraints imposed on tensors, whereas the number-state preserving tensors proposed in this manuscript are not required to be isometric. Similarly, configuration causal cones arise only in networks that preserve number states, and are thus ill-defined for generic MERA. [Note that it is, however, possible to have networks with tensors that are both simultaneously isometric and number-state preserving, see Sect. \ref{sect:A} of the Appendix for further discussion]. Despite the difference in the origins of these two forms of causal cone, it is not a fluke that they were exactly equivalent in the previous example. It can be understood that the configuration causal cones in any number-state preserving tensor network are always equivalent to the causal cones found in an isometric tensor network of the same geometry, provided that the index orientations (specifying incoming and outgoing indices) match between the networks. Given this equivalence, we will henceforth drop the distinction between the two definitions, such that the term `causal cone' can refer to either definition.

The process of evaluating the configuration space for a region $B$ of three sites from a binary MERA is depicted in Fig. \ref{fig:CausalReduce}(a). This evaluation can be formulated as a sequence of contractions that each `lower' the configuration space through the causal cone,
\begin{equation}
\ldots \ket{\phi ^B_{[2]}} \to  \ket{\phi ^B_{[1]}} \to  \ket{\phi ^B_{[0]}}
\end{equation}
where bracketed subscripts denote configuration spaces at different depths within the network. Each of the lowering contractions is implemented by one of two geometrically different lowering operators, depicted in Fig. \ref{fig:CausalReduce}(b), which are the direct analogues to the descending superoperators\cite{Alg1} used in the evaluation of density matrices from isometric MERA. 

In our example using a binary MERA, the cost of evaluating  $\ket{\phi _B^{[0]}}$ for a region $B$ of three contiguous sites scales at most linearly with the network depth, since the form of the lowering operators are self-similar at all depths. In a general (number-state) preserving network the computational cost of evaluating configuration spaces will be related to the causal structure of the network: the leading order cost will scale exponentially with maximum width of the causal cones. Thus it is apparent that not all number-state preserving tensor networks can be efficiently evaluated for local information (characterized by the configuration space $\ket{\phi _B}$); only those for which the maximum causal width is not too large. However, since MERA are precisely designed to have bounded causal width (i.e. the causal width never spreads beyond some small number of sites), it follows that number-state preserving versions of MERA networks precisely fall within the class of networks that can be efficiently evaluated.

%%%%%%%%%%%%%%%%%%%%%%%%%%%%%%%%%%%%%%
%%%%%%%%%%%%%%%%%%%%%%%%%%%%%%%%%%%%%%
\begin{figure}[!t!b]
\begin{center}
\includegraphics[width=8.5cm]{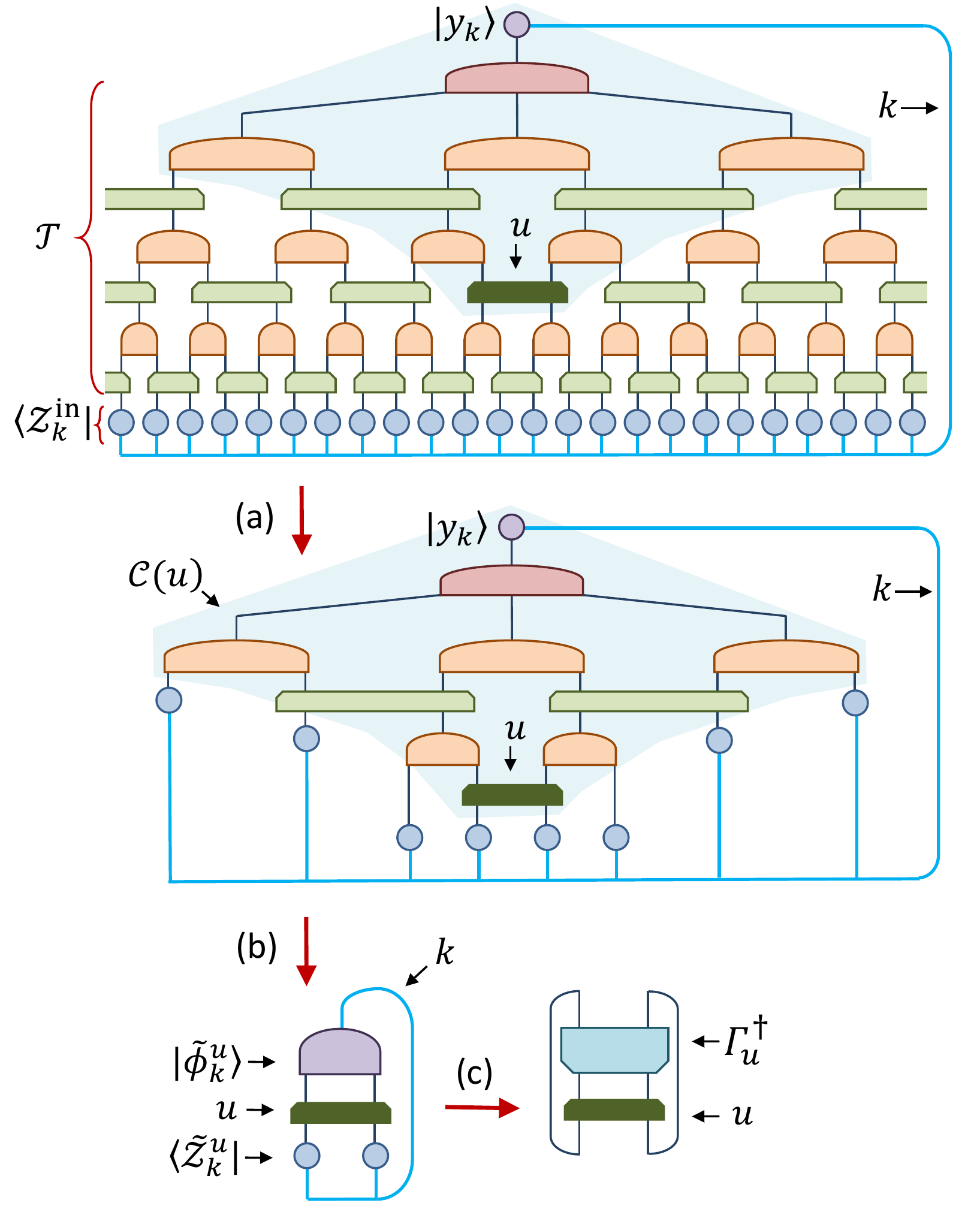}
\caption{The sequence of steps used to evaluate the environment $\Gamma_u$ of the shaded tensor $u$. (a) The initial state $\bra{{\Z}_k^\textrm{in}}$ is transformed through the network to form a new number state on the boundary of the causal cone $\C(u)$ associated to $u$. (b) The configuration space $\ket{\tilde \phi_k^u}$, defined on the output indices of $u$, is computed through use of the left/right lowering operators, as in Fig. \ref{fig:CausalReduce}. (c) The environment $\Gamma$ is formed by taking the outer product of the configuration space $\ket{\tilde \phi_k}$ with the state $\bra{{\tilde \Z}_k^u}$ defined the input of $u$, summing over all training samples $k$, see also Eq. \ref{eq:env}.} 
\label{fig:EnvEval}
\end{center}
\end{figure}
%%%%%%%%%%%%%%%%%%%%%%%%%%%%%%%%%%%%%%%%
%%%%%%%%%%%%%%%%%%%%%%%%%%%%%%%%%%%%%%%%

Given that the evaluation of configuration spaces has been understood, we now turn to the task of building the environment $\Gamma_u$ associated to tensor $u$, as depicted in Fig. \ref{fig:EnvEval}, which is accomplished as follows. First we lift the initial number state $\ket{\Z_k^\textrm{in}}$ to a new number state $\ket{\tilde{\Z}_k}$ that lives on the boundary of causal cone $\C(u)$ associated to tensor $u$, as depicted in Fig. \ref{fig:EnvEval}(a). Then we compute the configuration space $\bra{\tilde{\phi}_k^u}$ defined on the output indices of tensor $u$, as depicted in Fig. \ref{fig:EnvEval}(b). Then the environment $\Gamma_u$ is given by taking the outer product of the configuration space $\bra{\tilde \phi_k^u}$ with the piece of the state $\ket{\tilde{\Z}_k}$ supported on the input indices of $u$, denoted $\ket{{\tilde \Z}_k^u}$, while summing over all training samples $k$,
\begin{equation}
{\Gamma _u} = \sum\limits_k {\left| {{{\tilde {\Z}}_k^u}} \right\rangle \left\langle {{{\tilde \phi }_k^u}} \right|}, \label{eq:env}
\end{equation}
see also Fig. \ref{fig:EnvEval}(c).

\section{Benchmark results} \label{sect:bench}
In this section we present benchmark results for how number-state preserving tensor networks perform as classifiers in some simple problems. The goal here is to establish the feasibility of our proposal, rather than to establish performance for challenging real-world tasks, which will be considered in future work. In particular we demonstrate (i) that the proposed optimization algorithms can efficiently and reliably train the networks under consideration, and (ii) that number-preserving networks perform comparably well to unrestricted networks for classification tasks. 

\subsection{Parity classification} \label{sect:parity}
\begin{figure}[!t!b]
\begin{center}
\includegraphics[width=8.5cm]{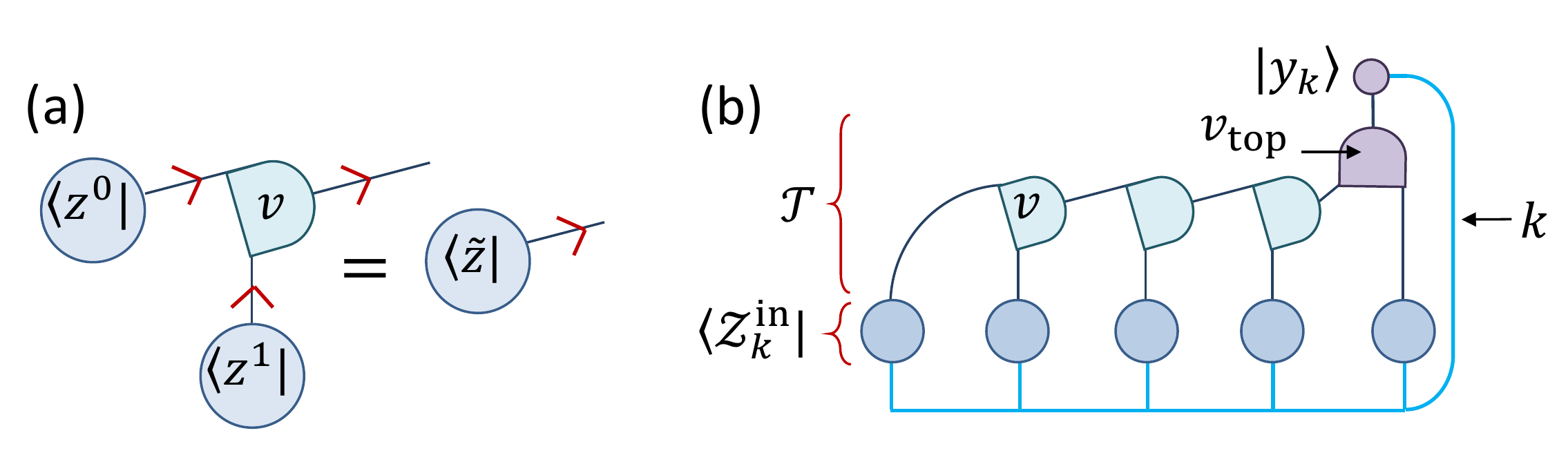}
\caption{(a) Tensor $v$ is a number-state preserving tensor mapping from two indices to a single index. (b) An MPS network $\T$ is built from tensors $v$ that preserve number-states when mapping from left-to-right. The MPS is trained as a classifier by maximizing the scalar product $\sum\nolimits_k {\left\langle {{\Z_k^\textrm{in}}} \right|\T\left| {y_k} \right\rangle }$.} 
\label{fig:MPS}
\end{center}
\end{figure}

For this first test, we benchmark the performance of a number-state preserving MPS for classifying the parity of binary strings. Here each test sample is a length-$N$ binary vector ${\vec \Z_k} = [0,0,1,0,1,\ldots]$, which is labeled $y_k \in \{0,1\}$ according to its parity. The MPS that we use is depicted in Fig. \ref{fig:MPS}, and is built from tensors that are number-state preserving only when acting from left-to-right. In this problem, we are free to choose the length $N$ of the binary strings as well as the number $n_\textrm{samp.}$ of training samples to use (as these can be randomly generated). We also have  two hyper-parameters associated to our method: the maximal bond dimension $\chi_\textrm{max}$ of the MPS and the parameter $\alpha$ from Eq. \ref{eq:prob} that controls the amount of randomness in the optimization. For each set of parameters investigated we performed 100 trial runs, each run starting with a randomly generated training set and a randomly initialized MPS, and then performed at no more than 100 optimization sweeps in each trial. The most computationally demanding trials (which consisted of: a length $N=20$ chain, $n_\textrm{samp.} = 20000$ training samples, a bond dimension of $\chi_\textrm{max}=10$, and 100 optimization sweeps) each took about 5 secs to run on a single 3 GHz desktop CPU. At the end of each trial we also test the generalization error of the MPS classifier by evaluating its accuracy in classifying the parity of all possible $2^N$ binary strings.  

A summary of the results from a large number of trials is presented in Tab. \ref{tab:MPSbench}. For binary strings of length $N=16$ and $N=20$ we used $1300$ and $20000$ training samples respectively; these numbers were chosen as they represent about $2\%$ of all possible binary strings in each case (of which there are $2^N$ in total). The randomness parameter was fixed at $\alpha=1$ for $N=16$ and $\alpha = 5$ for $N=20$ length chains; these values were determined as adequate through small amount of experimentation (and are probably not those which would give optimal performance). Somewhat surprisingly, we found that each trial would produce only one of two outcomes: (i) the optimization would fail completely, achieving only slightly over $50\%$ classification accuracy on the set of all binary strings, or (ii) would converge to a perfect parity classifier, with $100\%$ classification accuracy for all length-$N$ binary strings. From Tab. \ref{tab:MPSbench} we see the proportion $n_\textrm{perfect}$ of perfect classifiers obtained increases dramatically as the bond dimension $\chi_\textrm{max}$ was increased, reaching $96/100$ for $N=20$ and $\chi_\textrm{max}=10$. This is expected, as networks with more degrees of freedom are less likely to be trapped in local minima. We found that the likelihood of obtaining a perfect classifier was also greatly improved when using a larger number of training samples, although do not provide this data here. In a recent work by Stokes and Terilla\cite{Parity1} standard (unrestricted) MPS were also trained to classify the parity of binary strings, and produced comparable results for similar strings lengths and training set sizes. This is a good indication that, for this classification problem, number-state preserving MPS are as powerful as unrestricted MPS. 

%%%%%%%%%%%%%%%%%%%%%%%%%%%%%%%%%%%%%%
%%%%%%%%%%%%%%%%%%%%%%%%%%%%%%%%%%%%%%
%%%%%%%%%%%%%%%%%%%%%%%%%%%%%%%%%%%%%%%%
%%%%%%%%%%%%%%%%%%%%%%%%%%%%%%%%%%%%%%%%

%%%%%%%%%%%%%%%%%%%%%%%%%%%%%%%%%%%%%%%%
%%%%%%%%%%%%%%%%%%%%%%%%%%%%%%%%%%%%%%%%
\begin{table}[!b!t]
\centering
%\begin{subtable}{0.48\textwidth}
Parity Classification:
\begin{tabular}{|c|c|c|c|c|c|}
\hline
$\;\;\; N \;\;\;$ & $\;\; n_\textrm{samp} \;\; $ & $\chi_\textrm{max}$ & $\;\;\;\; \alpha \;\;\;\;$ & $\; \; n_\textrm{perfect} \; \; $ & $n_\textrm{sweeps}$ \\ \hline
16      & 1300    & 4   & 1   & 38/100    & 31          \\ 
16      & 1300    & 6   & 1   & 63/100    & 28          \\ 
16      & 1300    & 10  & 1   & 93/100    & 25          \\ 
20      & 20000   & 4   & 5      & 34/100    & 26          \\ 
20      & 20000   & 6   & 5      & 63/100    & 21          \\ 
20      & 20000   & 10  & 5      & 96/100    & 27          \\ \hline
\end{tabular}

\medskip
\noindent
Division-by-7 Classification:

\begin{tabular}{|c|c|c|c|c|c|}
\hline
$\;\;\; N \;\;\;$ & $\;\; n_\textrm{samp} \;\; $ & $\chi_\textrm{max}$ & $\;\;\;\; \alpha \;\;\;\;$ & $\; \; n_\textrm{perfect} \; \;$ & $n_\textrm{sweeps}$ \\ \hline
16      & 3000    & 9   & 1   & 92/100    & 43          \\ 
16      & 3000    & 12  & 1   & 100/100   & 36          \\ 
16      & 3000    & 16  & 1   & 98/100    & 29          \\ 
20      & 30000   & 9   & 5      & 75/100    & 56          \\ 
20      & 30000   & 12  & 5      & 88/100    & 44          \\ 
20      & 30000   & 16  & 5      & 96/100    & 26          \\ \hline
\end{tabular}
\caption{Summary of results for MPS applied to the parity classification (above) and division-by-7 classification (below). Parameters are as follows: $N$ is the length of binary strings classified, $n_\textrm{samp}$ is the number of samples in the training set, $\chi_\textrm{max}$ is the maximal MPS bond dimension, parameter $\alpha$ controls the randomness in the optimization as per Eq. \ref{eq:prob}, $n_\textrm{perfect}$ is the proportion of trial runs that yielded perfect ($100\%$ accuracy) classifiers, $n_\textrm{sweeps}$ is the average number of variational sweeps required to reach convergence.}  \label{tab:MPSbench}
\end{table}
%%%%%%%%%%%%%%%%%%%%%%%%%%%%%%%%%%%%%%%%
%%%%%%%%%%%%%%%%%%%%%%%%%%%%%%%%%%%%%%%%

\subsection{Division-by-7 classification} \label{sect:div}
For the second test we classify binary strings, interpreted as a base-2 representation of an integer, by their remainder under division by 7. We again use a number-state preserving MPS, employing the same set-up as used for the parity classification considered previously. A key difference here is that the samples now take one of seven different labels, $y_k \in \{0,1,2,3,4,5,6\}$.

A summary of the results from these trials is presented in Tab. \ref{tab:MPSbench}. For binary strings of length $N=16$ and $N=20$ we used $3000$ and $30000$ training samples respectively; although this was more than was used for the parity classification it is still less than $5\%$ of the possible binary strings. Similar to the parity benchmark, we here found that each trial would either fail completely, producing no better than a random results, or would converge to a perfect division classifier, with $100\%$ classification accuracy for all length-$N$ binary strings. As with the parity benchmark, it is seen that the proportion of perfect classifiers obtained increases steadily with the bond dimension $\chi_\textrm{max}$. However, this problem required larger dimensions $\chi_\textrm{max}$ than used for the parity benchmark, which is expected since here we have many more classification categories. 

%%%%%%%%%%%%%%%%%%%%%%%%%%%%%%%%%%%%%%
%%%%%%%%%%%%%%%%%%%%%%%%%%%%%%%%%%%%%%
\begin{figure}[!t!b]
\begin{center}
\includegraphics[width=8.5cm]{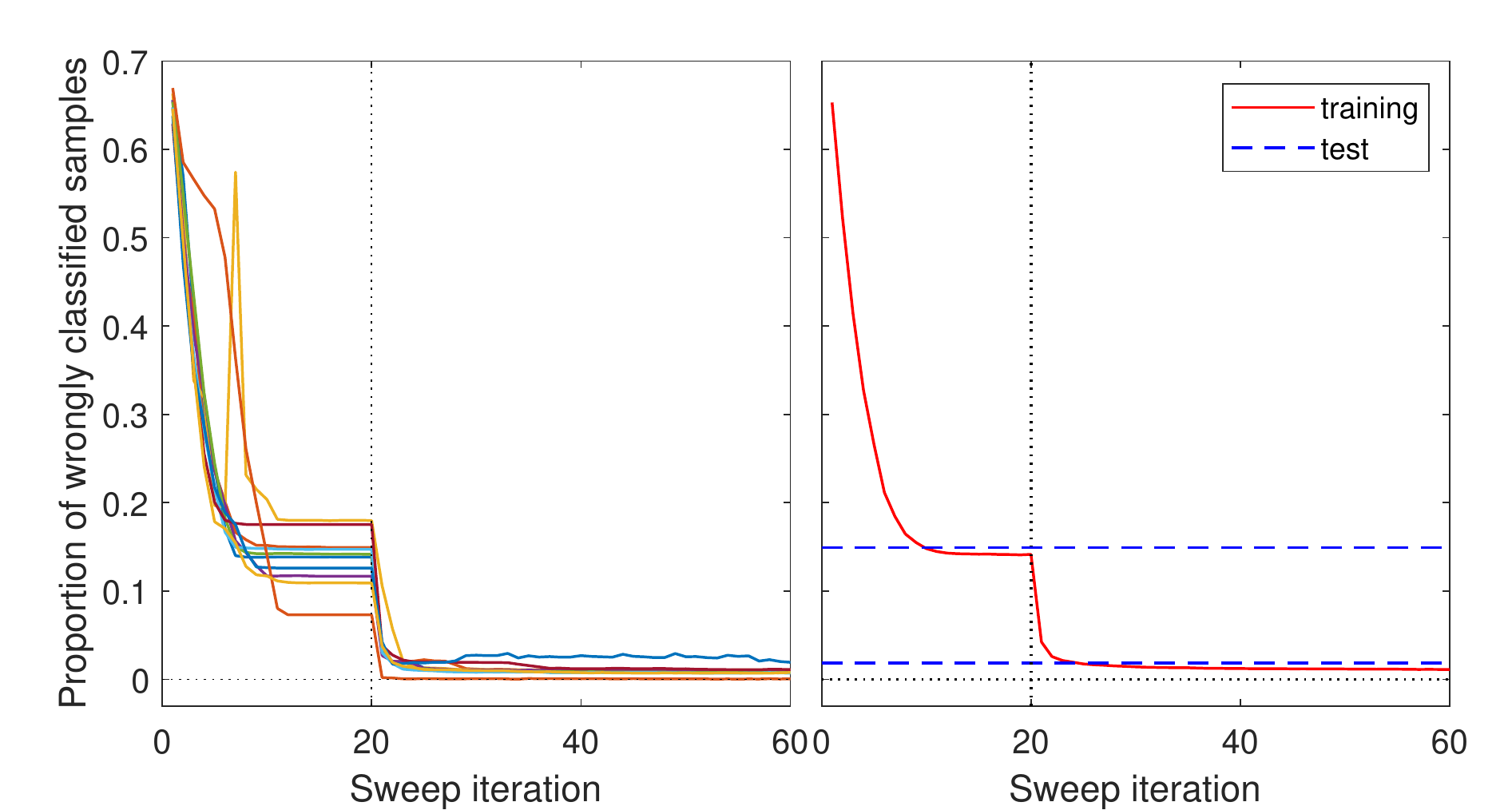}
\caption{(left) Results of training TTN and MERA for the height classification problem, displaying how much of training set is wrongly classified as a function of the number of optimization sweeps performed. The first 20 sweeps are performed while keeping trivial disentanglers $u$, such that underlying the network is a TTN, while the $u$ are then `switched on' for the remaining sweeps such that the network becomes a MERA. The figure displays results from 10 different trials, where each trial starts with a randomly generated training set and randomly initialized network. (right) Average results of the training data from 100 trial runs (after discarding the 10 worst trials). Dashed lines show the average generalization error computed from applying the trained TTN and MERA applied to a randomly generated test set. For TTN we get average training/test errors of $14.15\%$ and $14.91\%$, while for MERA we get average training/test errors of $1.13\%$ and $1.86\%$.} 
\label{fig:MERAdata}
\end{center}
\end{figure}
%%%%%%%%%%%%%%%%%%%%%%%%%%%%%%%%%%%%%%%%
%%%%%%%%%%%%%%%%%%%%%%%%%%%%%%%%%%%%%%%%

\subsection{Height classification} \label{sect:MERA}
The final test problem that we consider, which we refer to as height classification, takes length-$N$ strings of integers from the set $z\in\{-1,0,1\}$ and classifies them with labels $y_k \in \{0,1,2\}$ depending on whether the sum (under regular addition) of the integers is positive, zero or negative, respectively. We test the effectiveness of both number-state preserving binary TTN and binary MERA as classifiers for this problem, working with strings of length $N=24$. A binary MERA of the form depicted in Fig. \ref{fig:NSnetwork}(b) is used, and is compared with the binary TTN that would result from restricting to trivial disentanglers $u$ throughout the MERA network. Given that the problem is translation-invariant, we imposed that all tensors within a network layer are identical. In terms of the optimization, this is achieved by updating using the average single-tensor environment from all equivalent tensors within a network layer. We found that the injection of randomness into the optimization was unnecessary, possibly due to the imposition of translational invariance, such that the randomness parameter $\alpha$ from Eq. \ref{eq:prob} could be set at $\alpha=0$. This left the bond dimension of the networks as the only hyper-parameter in the calculation, which was fixed at maximum dimension $\chi_\textrm{max} = 9$. 

The benchmark results are displayed in Fig. \ref{fig:MERAdata}, and consisted of 100 trials, each trial starting from $12000$ randomly generated training samples (with $4000$ samples from each label category) and a randomly initialized network. Rather than running separate TTN and MERA trials they were instead combined: the first 20 sweeps were performed with trivial disentanglers $u$, such that underlying the network was a TTN, the $u$ were then `switched on' for the remaining 40 sweeps such that the network became a MERA. At the conclusion of each trial, the generalization error was estimated by applying the trained classifiers to a randomly generated test set of the same size as the training set. Most of the trials converged smoothly, with the proportion of wrongly identified testing samples decreasing monotonically with optimization, although about 5 trials failed to properly converge (yielding classifiers with greater than $30\%$ error). Discarding the worst 10 trials from consideration, of the 90 remaining trials the TTN gave average training/test errors of $14.15\%$ and $14.91\%$, while MERA gave substantially reduced average training/test errors of $1.13\%$ and $1.86\%$. These results clearly demonstrate the extra representation power endowed through use of the disentanglers $u$ in MERA. Impressive is that both networks generalized well, with only relatively small differences between test and training accuracies, despite being trained on less than $5\times 10^{-6}$ percent of the possible $3^{24}$ training samples.
  
\section{Conclusions} \label{sect:conclusion}
We have proposed the class of number-state preserving tensor networks for use as classifiers in supervised learning tasks and have shown that a large class of these networks, specifically those with bounded causal structure, are efficiently trainable for large problems. In particular we have described a training algorithm that, for any chosen tensor in the network under consideration, exactly identifies the optimal tensor for that location (i.e. that which maximizes the number of correctly classified training samples), all with cost that scales only linearly in number of training samples. Importantly, the class of efficiently trainable number-state preserving networks includes realizations of sophisticated networks such as MERA, which would otherwise be computationally intractable. As such, we believe this could be the first computationally viable proposal which would allow MERA, close tensor network analogues to convolutional neural networks, to be applied as classifiers for challenging tasks such as image recognition. This remains an interesting direction for future research. 

Although number-state preserving tensors represent a highly restricted class of tensor, the preliminary results of Sect. \ref{sect:bench} are encouraging that this class is sufficient when applying tensor networks as classifiers for learning problems as outlined in Sect. \ref{sect:supervised}. It still remains to be seen whether number-state preserving tensor networks are as powerful as generic tensors networks for these tasks; this question requires further theoretical and numerical investigation. However it is relatively easy to understand that, in the limit of large bond dimension, a number-state preserving tensor network could in principle achieve $100\%$ accuracy on any training problem outlined in Sect. \ref{sect:supervised}. The reasoning follows similarly to the argument that a generic tensor network can represent an arbitrary quantum state in the limit of large bond dimension. Consider, for instance, the MERA depicted in Fig. \ref{fig:NSnetwork}(b). One could increase the bond dimension of indices within the network until the output index of each $w$ tensor matches the product of its input dimensions, in which case each $w$ could be fixed as a trivial identity tensor when viewed as an input-output matrix. In this scenario, the top tensor $w_\textrm{top}$ could implement an arbitrary classifier that would perfectly map every training sample to its designated label, regardless of the training data given. 

%Require going to 2D networks such as 2D MERA. MERA already proven much more effective than tree for 1D problem, effect is much more pronouned for 2D. 

A major difficulty with the use of MERA in $D=2$ or higher spatial dimensions\cite{MERA2,MERA3} is their high scaling of computational cost with bond dimension $\chi$. However, there is reason to be more optimistic for their application as classifiers. The cost of contracting an isometric metric MERA for a density matrix, necessary for its optimization towards the ground state of a local Hamiltonian, is related to the size of the maximum causal width of the network. For instance, the most efficient known $2D$ isometric MERA\cite{MERA3} has a causal width of $2\times 2$ sites, such that the density matrices within the causal cone have 8 indices. The cost of computing these density matrices can be shown to scale at most as $O(\chi^{16})$. However, while a number-state preserving version of this $2D$ MERA would also have a causal width of $2\times 2$ sites, the relevant configuration space $\ket{\psi}$ within the causal cone would only have 4 indices (which follows as the density matrix involves both the \emph{bra} and the \emph{ket} state, whereas the configuration space only involves the \emph{ket}). Thus the cost of optimizing a number-state preserving version of this $2D$ MERA, where the key step is the evaluation of configuration spaces, will scale roughly as $O(\chi^{8})$ (i.e. the square-root of the cost of optimizing an isometric MERA for a quantum ground state). This square-root reduction in cost scaling as a function of bond dimension $\chi$ from isometric to number-state preserving networks will hold in general, such that number-state preserving networks could realize much larger bond dimensions given a fixed computational budget. This advantage is somewhat mitigated by the fact that the cost of optimizing a number-state preserving network comes with a factor ${n_{samp}}$ related to the size of the training set, which could be very large. However, it would also be straight-forward to parallelize the evaluation of environments over the samples.

Although the main text of this manuscript focused on number-state preserving versions of MERA, many other forms of hierarchical network could also be of useful as classifiers as discussed further in Sect. \ref{sect:B} of the Appendix. In particular the network of Fig. \ref{fig:OtherMERA}, which does not have an isometric counterpart, seems to be the closest tensor network analogue to a convolutional neural network. Rather than disentanglers, this network uses $\delta$-function tensors to effectively allow neighboring $w$ tensors to `read' from the same boundary sites, mirroring the overlap of feature maps arising in a convolution (and similar to the generalized networks recently proposed in Ref. \onlinecite{Glas18b}). It would be interesting to compare the effectiveness of this structure versus a traditional MERA, which will be considered in future work.

%%%% Acknowledgements
The author thanks Miles Stoudenmire and John Terilla for useful discussions and comments. This research was supported in part by the National Science Foundation under Grant No. NSF PHY-1748958.

%%%%%%%%%%%%%%%%%%%%%%%%%%%%%%%%%%%%%%%%%%%%%%%%%%%%%%%%%%%

\appendix
\newpage
\phantom{hello}
\newpage

\section{Classes of number-state preserving tensors} \label{sect:A}
\begin{figure}[!t!b]
\begin{center}
\includegraphics[width=8.5cm]{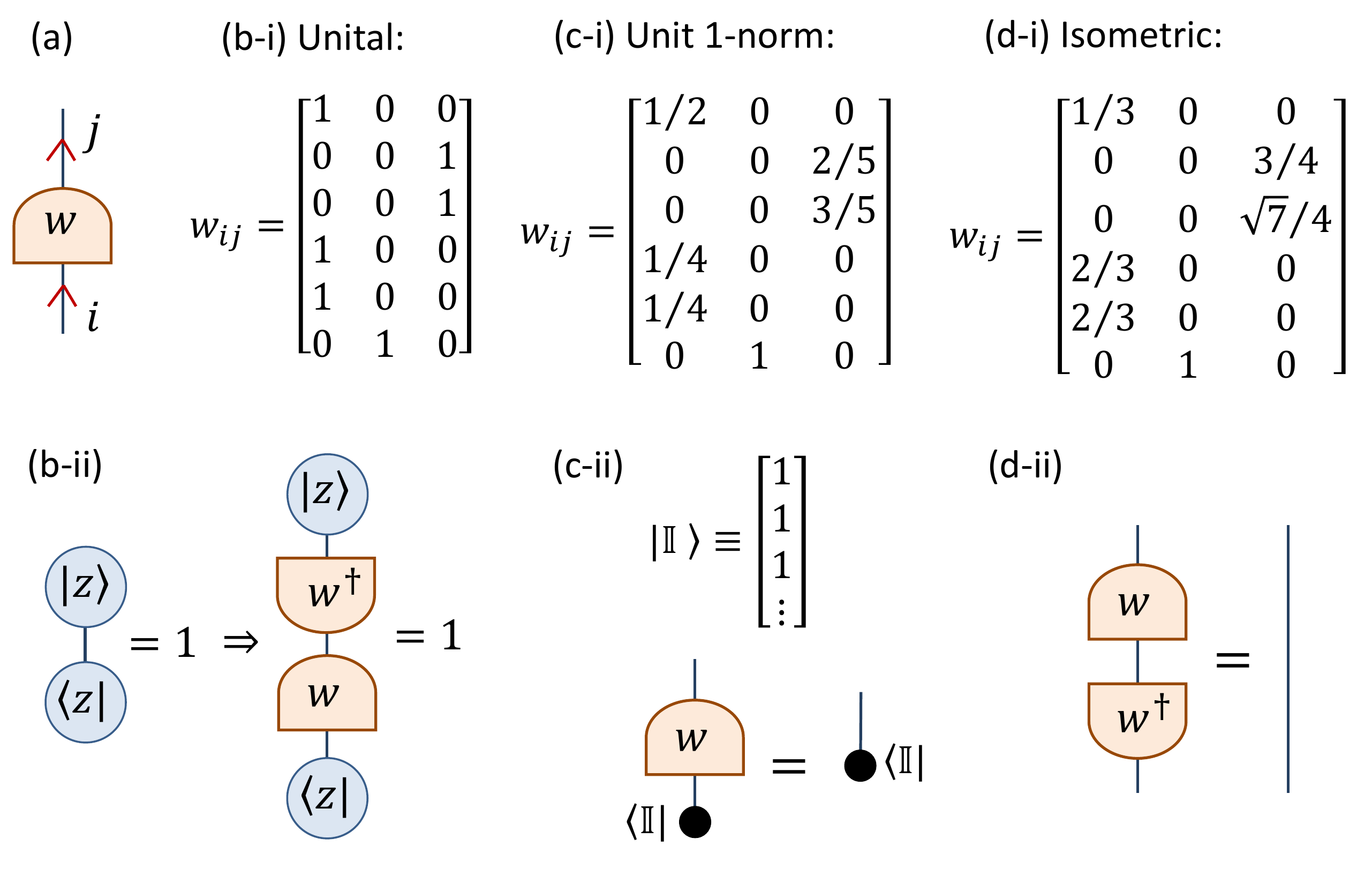}
\caption{(a) Tensor $w$ is assumed to be a number-state preserving tensor with a single input and a single output index. (b) An example of a unital tensor $w$, which preserves the norm of number states under transformations, i.e. such that $\bra{z} w w^\dag \ket{z} = 1$ for any normalized number-state $\ket{z}$. (c)  An example of a tensor $w$ with unit 1-norm, which transforms an equal superposition vector $\left| \mathbb{I} \right\rangle  = {[1,1,1,1, \cdots ]^\dag }$ into another equal superposition vector. (d) An example of an isometric tensor $w$, which annihilates to the identity $I$ under contraction with its conjugate tensor, ${w^\dag }w = I$.} 
\label{fig:MoreNorm}
\end{center}
\end{figure}

The numerical examples considered in the main text trained classifiers using tensor networks built from \emph{unital} number-state preserving tensors, where all tensor elements are either zero or the unit element. Using unital tensors has the advantage that they preserve the norm of number states under transformation, see Fig. \ref{fig:MoreNorm}(b), simplifying the cost function for identifying the number of correctly classified training samples. However, there are good reasons why one might want also want to consider non-unital tensors. A classifier built with unital tensors only gives a binary result $\{0,1\}$ for whether a test state belongs to a specified category. In practice it may be preferable to obtain a continuous parameter in the range $p\in[0,1]$ that indicates the likelihood of a test state belonging to the specified category, which could be achieved using number-state preserving tensors with arbitrary real entries. We now consider two potentially useful forms of non-unital tensors that are still number-state preserving.

%%%%%%%%%%%%%%%%%%%%%%%%%%%%%%%%%%%%%%
%%%%%%%%%%%%%%%%%%%%%%%%%%%%%%%%%%%%%%
%%%%%%%%%%%%%%%%%%%%%%%%%%%%%%%%%%%%%%%%
%%%%%%%%%%%%%%%%%%%%%%%%%%%%%%%%%%%%%%%%

A useful class of number-state preserving tensor to consider are those with unit 1-norm, as per the example of Fig. \ref{fig:MoreNorm}(c). These are tensors that, when expressed as an input-output matrix, have columns that sum to unity. This property implies that these tensors transform an equal superposition vector $\left| \mathbb{I} \right\rangle  = {[1,1,1,1, \cdots ]^\dag }$ on their input into an equal superposition vector on their output. It follows that a tensor network built from these will have a 1-norm of unity. The restriction to tensors with of this normalization also has the advantage in that it allows marginal probability distributions to be evaluated from number-state preserving networks, which may otherwise not be feasible. Assume that we wish to evaluate from a tensor network classifier the weighted set of permissible configurations for some region $B$ while knowing nothing of the state on the complimentary region $A$ of the problem space, which we call the marginal distribution for region $B$. We can compute the marginal distribution by repeating the calculation from Sect. \ref{sect:enviro} for the configuration space for $B$, but instead setting the state $\ket{{\Z^A}}$ on the compliment as the equal superposition,
\begin{equation}
\left| {{\Z^A}} \right\rangle  = \left| \mathbb{I} \right\rangle \left| \mathbb{I} \right\rangle \left| \mathbb{I} \right\rangle \left| \mathbb{I} \right\rangle  \ldots.
\end{equation}
This evaluation can be performed efficiently, since tensors with unit 1-norm map the superposition vector $\ket{\mathbb{I}}$ trivially to itself.

In certain cases it is also possible to restrict tensors to be both simultaneously number-state preserving and isometric, see Fig. \ref{fig:MoreNorm}(d) for an example. This is only possible if the product of the incoming dimensions is greater than or equal to the product of the outgoing dimensions, which is necessary for the isometric character. A network built from these tensors would inherit both the efficient evaluation of reduced density matrices, characteristic to isometric networks, and the efficient evaluation of configuration spaces, characteristic to number-state preserving networks. In addition, restricting to isometric tensors ensures that the 2-norm of a network is unity.

%%%%%%%%%%%%%%%%%%%%%%%%%%%%%%%%%%%%%%
%%%%%%%%%%%%%%%%%%%%%%%%%%%%%%%%%%%%%%
\begin{figure}[!t!b]
\begin{center}
\includegraphics[width=8.0cm]{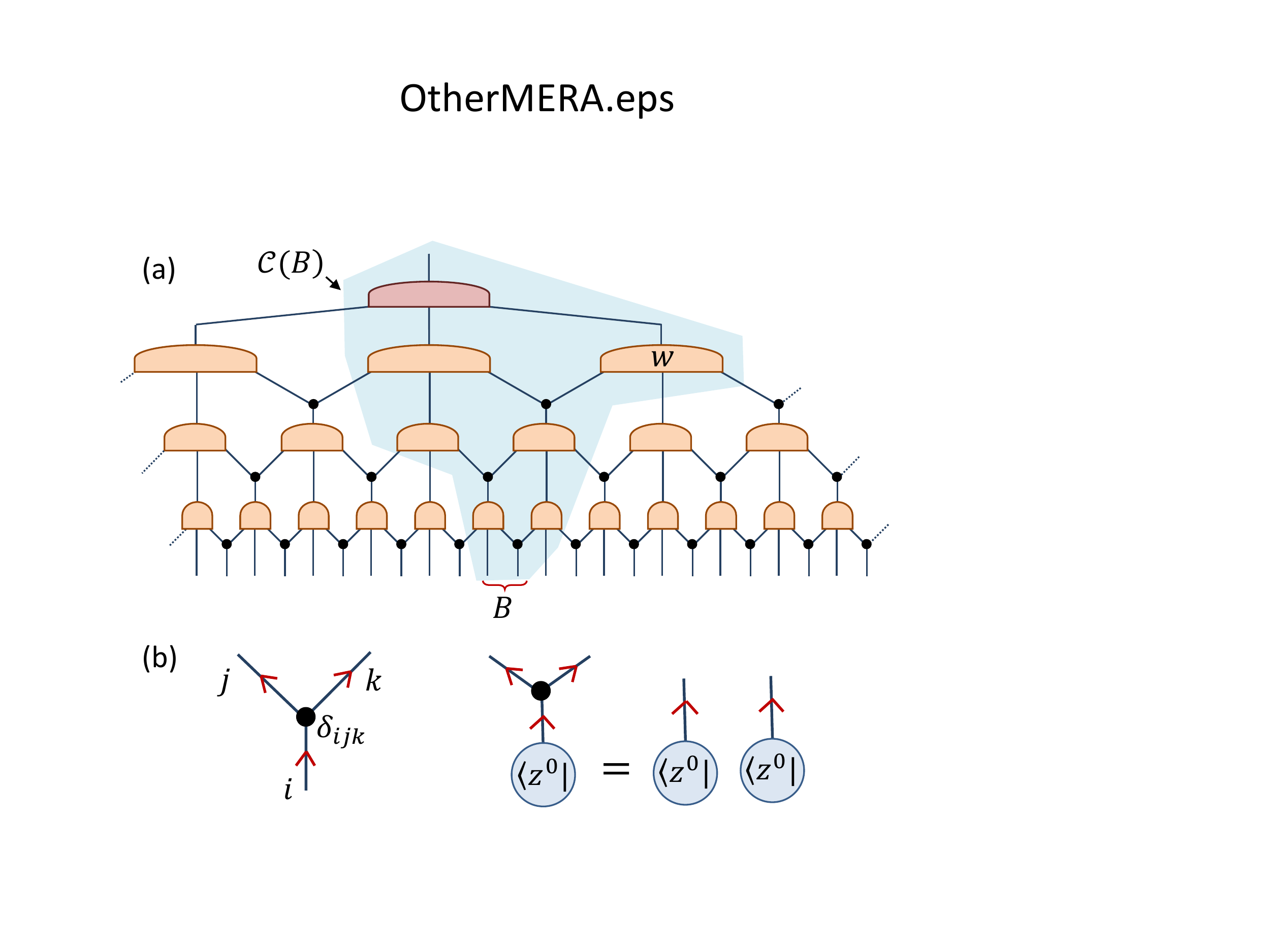}
\caption{(a) A hierarchical tensor network constructed mimic a convolutional neural network (CNN). (b) The black circles represent the $\delta$-function, which maps a number state into two copies of itself, thus two adjacent $w$ tensors are able to effectively `read' from the same lattice site. The causal cone $\C(B)$ of region $B$ is shaded, which has a bounded width of two sites.} 
\label{fig:OtherMERA}
\end{center}
\end{figure}
%%%%%%%%%%%%%%%%%%%%%%%%%%%%%%%%%%%%%%%%
%%%%%%%%%%%%%%%%%%%%%%%%%%%%%%%%%%%%%%%%

\section{Alternative hierarchical networks} \label{sect:B}
In main text we considered mainly number-state preserving versions of standard tensor networks (including MPS, TTN and MERA). However, many other forms of number-state preserving networks may also be useful as classifiers, some of which fall outside of what is permissible with isometric networks. In this appendix we give a few examples of more general networks and discuss where they may be useful.

Consider the example MERA-like network depicted in Fig. \ref{fig:OtherMERA}. Unlike a traditional MERA this network does not use disentanglers, instead using $\delta$-function tensors to effectively allow neighboring $w$ tensors to `read' from the same boundary sites, similar to the generalized networks recently proposed in Ref. \onlinecite{Glas18b}. Notice that this construction is not compatible with imposing an isometric character on the tensors. This network seems to be close analogue to a convolutional neural network, in that the $\delta$-function tensors mimic the overlapping feature maps arising in a convolution. The cost of optimizing this network for a supervised learning problem is seen to be cheaper than that of the binary MERA considered in the main text, since the causal cones here only have maximal width of two sites. Given this consideration, it will be interesting to see how the accuracy compares with binary MERA, which we leave for future work.  

%%%%%%%%%%%%%%%%%%%%%%%%%%%%%%%%%%%%%%
%%%%%%%%%%%%%%%%%%%%%%%%%%%%%%%%%%%%%%
\begin{figure}[!t!b]
\begin{center}
\includegraphics[width=8.0cm]{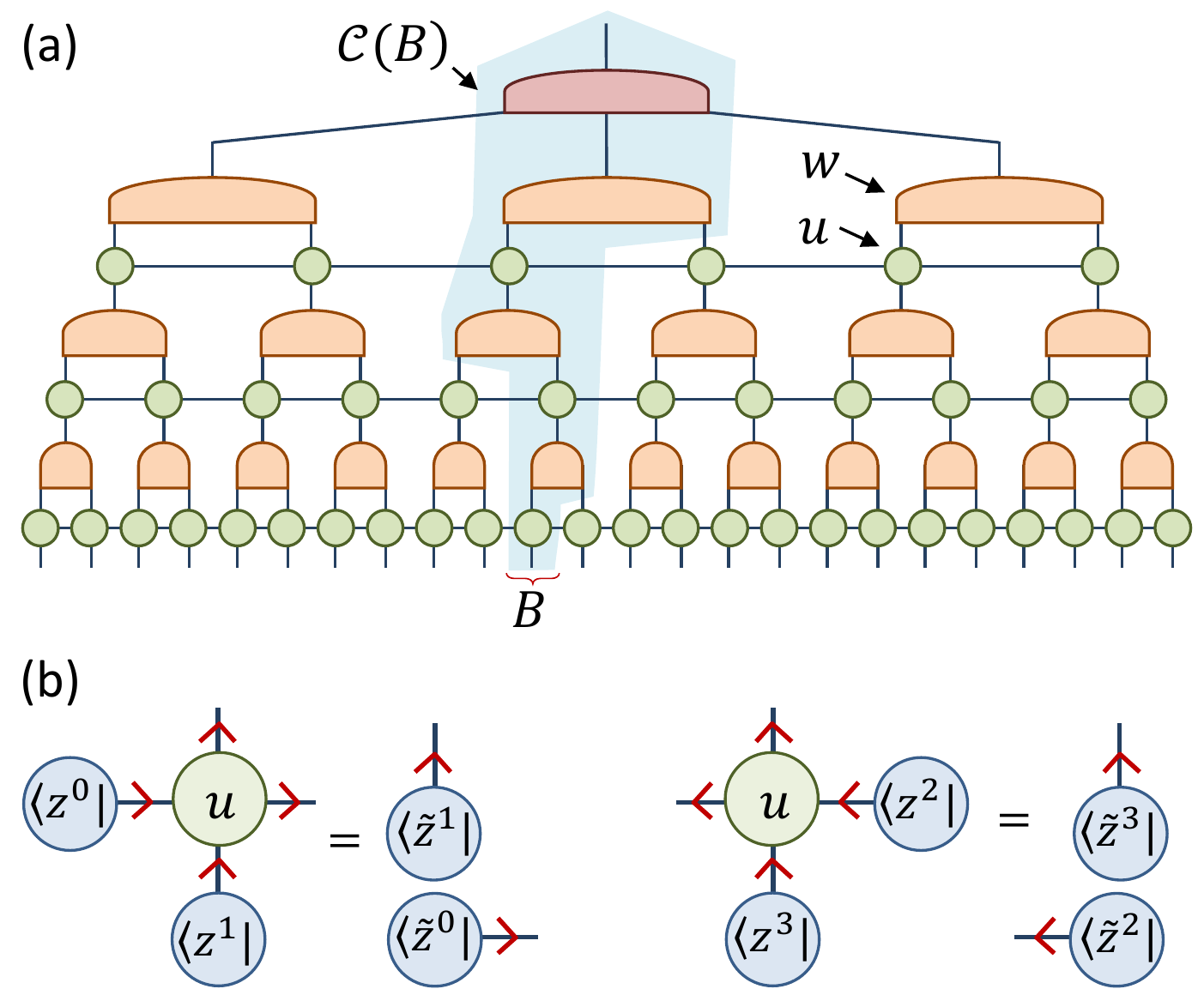}
\caption{(a) A hierarchical tensor network where the disentangling is accomplished via matrix product operators (MPOs) of four index tensors $u$. (b) In order for the network to have a bounded causal width, each $u$ must be simultaneously number-state preserving with respect to the two orientations pictured. } 
\label{fig:OtherMERA2}
\end{center}
\end{figure}
%%%%%%%%%%%%%%%%%%%%%%%%%%%%%%%%%%%%%%%%
%%%%%%%%%%%%%%%%%%%%%%%%%%%%%%%%%%%%%%%%

Another type of MERA-like network is depicted in Fig. \ref{fig:OtherMERA2}(a); this time accomplishing disentangling using matrix product operators (MPOs) rather than a product of local tensors. In order for this network to have bounded causal width, and thus be compatible with efficient optimization, it is necessary that the $u$ tensors are simultaneously number-state preserving with respect to two different orientations, as depicted in Fig. \ref{fig:OtherMERA2}(b). If this criteria is satisfied, then the network will possess a causal-width of only one site and thus will be extremely efficient to optimize. There is some evidence to suggest that disentangling using MPOs could be much more effective that disentangling using local operators as used in a standard MERA. In a recent work\cite{Motz2}, this form of number-state preserving tensor network with bond dimension $\chi=4$ was shown to exactly describe the ground state of the Motzkin spin chain\cite{Motz1}, which possesses a logarithmic scaling of entanglement entropy. In contrast it is known that a regular MERA network, with arbitrarily large but finite bond dimension, cannot provide an exact representation of this ground state.

\end{document}